\pdfoutput=1
\documentclass{ceurart}

\usepackage{amssymb}
\usepackage{amsmath}
\usepackage{amsfonts}
\usepackage{amscd}
\usepackage{graphicx}
\usepackage{adjustbox}
\usepackage{enumitem}
\usepackage{booktabs}
\usepackage{multicol}
\usepackage{siunitx}

\usepackage{float}
\usepackage{listings}
\usepackage{latexsym}
\usepackage{epsf}
\usepackage{url}

\usepackage{multirow}
\usepackage{caption}
\usepackage{subcaption}
\usepackage[noabbrev]{cleveref}
\usepackage{microtype}
\usepackage{mathtools}
\usepackage{fixmath}
\renewcommand{\vec}[1]{\mathbold{#1}}

\usepackage{verbatim}
\usepackage{makecell}

\usepackage{pgfplots}
\usepackage{pgfplotstable}
\usepgfplotslibrary{statistics}
\pgfplotsset{compat=1.18}

\newcommand{\tout}{\text{timeout}}

\newcommand{\limp}{\mathbin\rightarrow}

\makeatletter
\g@addto@macro\bfseries{\boldmath}
\makeatother

\newcommand{\CaDiCaL}{\textsc{CaDiCaL}}

\newcommand{\libexact}{\textsc{libexact}}
\newcommand{\MOLS}[2]{#1\operatorname{MOLS}(#2)}

\begin{document}

\title{Improving SAT Solvers on Orthogonal Latin Square Problems}

\author[1]{Aaron Barnoff}[email=barnoffa@uwindsor.ca]
\address[1]{School of Computer Science, University of Windsor, Canada}

\author[2]{Curtis Bright}[email=cbright@uwaterloo.ca]
\address[2]{School of Computer Science, University of Waterloo, Canada}

\begin{abstract}
Latin squares are $n\times n$ matrices containing $n$ symbols, where each symbol appears exactly once in each row and column. They were studied by Euler, later popularized through Sudoku, and remain a rich source of difficult combinatorial search problems. Two Latin squares are orthogonal mates if, when overlaid, no ordered pair of symbols repeats. Pairs of orthogonal Latin squares exist for every order except 2 and 6, but finding orthogonal Latin squares computationally can be challenging.

Satisfiability (SAT) solvers are strong at combinatorial search and have been used to resolve a number of various kinds of orthogonal Latin square problems. On the other hand, SAT solvers lack domain knowledge about Latin squares, such as the Euler--Parker algorithm for orthogonal mate construction. In this paper, we propose a hybrid method combining a SAT solver with the Euler--Parker algorithm (implemented using a Diophantine system solver) and show that the resulting solver is effective at finding certain kinds of orthogonal Latin squares.
For example, certain pairs of $10\times10$ orthogonal Latin squares whose existence was unknown for over 25 years were recently found by Bright, Keita, and Stevens using a SAT solver.
The hardest cases
could not be solved by the SAT solver {\CaDiCaL} within seven days, but {\CaDiCaL} augmented with
an external Euler--Parker algorithm solves these cases in a median of around 5,100 seconds.
\end{abstract}


\copyrightyear{2026}
\copyrightclause{Copyright for this paper by its authors.
  Use permitted under Creative Commons License Attribution 4.0
  International (CC BY 4.0).}
\conference{11th International Workshop on Satisfiability Checking and Symbolic Computation, July 13, 2026, Oldenburg, Germany}

\maketitle

\section{Introduction} \label{sec:intro}

A Latin square of order $n$ is an $n \times n$ array on $n$ symbols in which each symbol occurs exactly once in each row and exactly once in each column (see Figure~\ref{fig:intro_a}).
Two Latin squares of the same order are \emph{orthogonal mates} if, when overlaid, every ordered pair of symbols occurs exactly once (see Figures~\ref{fig:intro_c}--\ref{fig:intro_d}).
Orthogonal Latin squares have a long history and are studied for their applications (e.g., to experimental design and coding theory) as well as
their relationship to other mathematical objects (e.g., $n-1$ mutually orthogonal Latin squares are equivalent to
a projective plane of order~$n$~\cite{Bose1938}).
Euler conjectured in 1782~\cite{euler1782recherches} that pairs of orthogonal Latin squares do not exist for orders~$n$ of the form $4k+2$.
The conjecture is trivial for $n=2$, but it took over a hundred years before Tarry~\cite{tarry1900probleme} proved the
next case of the conjecture for $n=6$ in 1900.
In the early 1900s, a number of mathematicians attempted to settle Euler's conjecture,
resulting in three independent publications laying claim to proving Euler's conjecture~\cite{peterson1902les,wernicke1910problem,macneish1922euler}.
It was thus quite surprising when Bose, Shrikhande, and Parker~\cite{BoseShrikhande1959,Parker1959,Bose} in 1959--1960
showed by construction that in fact orthogonal Latin squares \textbf{do} exist for every order $n$ with the exception of $2$ and $6$
and thus the ``proofs'' of Euler's conjecture in the literature were flawed.

Bose, Shrikhande, and Parker's resolution of Euler's conjecture was completed
without reliance on computers, although electronic computers had been
used to search for counterexamples to Euler's conjecture as early as 1953~\cite{Federer1971}.
During the 10th Symposium in Applied Mathematics in 1958, Paige and Tompkins~\cite{Paige1960} presented
results of a computer search for a pair of $10\times10$ orthogonal Latin squares.
They were unsuccessful in finding such a pair, and
estimated their program would need around $5\cdot10^{11}$ hours to determine if a given Latin square of order 10 has
an orthogonal mate.  On the other hand, in 1959, Parker~\cite{parker1959computer} had also written a
computer program that only needed about an hour to find all orthogonal mates of a given Latin square of order 10.
The primary reason why Parker's program was so much faster than Paige and Tompkins' program was because
Parker's program exploited a clever observation of Euler---that a Latin square of order $n$ has an orthogonal mate if and only if
it can be decomposed into $n$ disjoint (i.e., non-overlapping) transversals.  A \emph{transversal} in a Latin square of order~$n$ is a collection of
$n$ cells in the square with exactly one cell from each row, exactly one cell from each column, and containing
exactly one occurrence of each symbol.  For example, in Figure~\ref{fig:intro_a} there are five transversals
highlighted (the cells of each distinct colour form a distinct transversal).

\begin{figure}[t]
\centering

\begin{subfigure}[t]{0.3\textwidth}
    \centering
\definecolor{cellred}{RGB}{229,115,115}
\definecolor{cellblue}{RGB}{100,181,246}
\definecolor{cellpurple}{RGB}{186,104,200}
\definecolor{cellyellow}{RGB}{255,241,118}
\definecolor{cellgreen}{RGB}{129,199,132}

\newcommand{\FillCell}[3]{%
  \pgfmathtruncatemacro{\yy}{4-#2}
  \ifcase#3\relax
    \path[fill=cellred,draw=none] (#1,\yy) rectangle ++(1,1);
  \or
    \path[fill=cellblue,draw=none] (#1,\yy) rectangle ++(1,1);
  \or
    \path[fill=cellpurple,draw=none] (#1,\yy) rectangle ++(1,1);
  \or
    \path[fill=cellyellow,draw=none] (#1,\yy) rectangle ++(1,1);
  \or
    \path[fill=cellgreen,draw=none] (#1,\yy) rectangle ++(1,1);
  \fi
}

\newcommand{\PutDigit}[3]{%
  \pgfmathtruncatemacro{\yy}{4-#2}
  \node[inner sep=0pt,font=\fontsize{18}{18}\selectfont] at ({#1+0.5},{\yy+0.5}) {#3};
}

\newcommand{\PutPair}[3]{%
  \pgfmathtruncatemacro{\yy}{4-#2}
  \node[inner sep=0pt,font=\fontsize{15}{15}\selectfont] at ({#1+0.5},{\yy+0.5}) {#3};
}

\newcommand{\DrawGrid}{%
  \foreach \k in {0,1,2,3,4,5} {
    \draw[black, line width=0.4pt] (\k,0) -- (\k,5);
    \draw[black, line width=0.4pt] (0,\k) -- (5,\k);
  }
}

\begin{tikzpicture}[x=0.92cm,y=0.92cm]
  \foreach \c/\r/\clr in {
    0/0/0, 1/0/1, 2/0/2, 3/0/3, 4/0/4,
    0/1/3, 1/1/4, 2/1/0, 3/1/1, 4/1/2,
    0/2/1, 1/2/2, 2/2/3, 3/2/4, 4/2/0,
    0/3/4, 1/3/0, 2/3/1, 3/3/2, 4/3/3,
    0/4/2, 1/4/3, 2/4/4, 3/4/0, 4/4/1
  }{
    \FillCell{\c}{\r}{\clr}
  }

  \DrawGrid

  \foreach \c/\r/\txt in {
    0/0/0, 1/0/1, 2/0/2, 3/0/3, 4/0/4,
    0/1/1, 1/1/2, 2/1/3, 3/1/4, 4/1/0,
    0/2/2, 1/2/3, 2/2/4, 3/2/0, 4/2/1,
    0/3/3, 1/3/4, 2/3/0, 3/3/1, 4/3/2,
    0/4/4, 1/4/0, 2/4/1, 3/4/2, 4/4/3
  }{
    \PutDigit{\c}{\r}{\txt}
  }

\end{tikzpicture}     \caption{Original square}
    \label{fig:intro_a}
\end{subfigure}
\hfill
\begin{subfigure}[t]{0.3\textwidth}
    \centering
\definecolor{cellred}{RGB}{229,115,115}
\definecolor{cellblue}{RGB}{100,181,246}
\definecolor{cellpurple}{RGB}{186,104,200}
\definecolor{cellyellow}{RGB}{255,241,118}
\definecolor{cellgreen}{RGB}{129,199,132}

\newcommand{\FillCell}[3]{%
  \pgfmathtruncatemacro{\yy}{4-#2}
  \ifcase#3\relax
    \path[fill=cellred,draw=none] (#1,\yy) rectangle ++(1,1);
  \or
    \path[fill=cellblue,draw=none] (#1,\yy) rectangle ++(1,1);
  \or
    \path[fill=cellpurple,draw=none] (#1,\yy) rectangle ++(1,1);
  \or
    \path[fill=cellyellow,draw=none] (#1,\yy) rectangle ++(1,1);
  \or
    \path[fill=cellgreen,draw=none] (#1,\yy) rectangle ++(1,1);
  \fi
}

\newcommand{\PutDigit}[3]{%
  \pgfmathtruncatemacro{\yy}{4-#2}
  \node[inner sep=0pt,font=\fontsize{18}{18}\selectfont] at ({#1+0.5},{\yy+0.5}) {#3};
}

\newcommand{\PutPair}[3]{%
  \pgfmathtruncatemacro{\yy}{4-#2}
  \node[inner sep=0pt,font=\fontsize{15}{15}\selectfont] at ({#1+0.5},{\yy+0.5}) {#3};
}

\newcommand{\DrawGrid}{%
  \foreach \k in {0,1,2,3,4,5} {
    \draw[black, line width=0.4pt] (\k,0) -- (\k,5);
    \draw[black, line width=0.4pt] (0,\k) -- (5,\k);
  }
}

\begin{tikzpicture}[x=0.92cm,y=0.92cm]
  \foreach \c/\r/\clr in {
    0/0/0, 1/0/1, 2/0/2, 3/0/3, 4/0/4,
    0/1/3, 1/1/4, 2/1/0, 3/1/1, 4/1/2,
    0/2/1, 1/2/2, 2/2/3, 3/2/4, 4/2/0,
    0/3/4, 1/3/0, 2/3/1, 3/3/2, 4/3/3,
    0/4/2, 1/4/3, 2/4/4, 3/4/0, 4/4/1
  }{
    \FillCell{\c}{\r}{\clr}
  }

  \DrawGrid

  \foreach \c/\r/\txt in {
    0/0/0, 1/0/1, 2/0/4, 3/0/2, 4/0/3,
    0/1/2, 1/1/3, 2/1/0, 3/1/1, 4/1/4,
    0/2/1, 1/2/4, 2/2/2, 3/2/3, 4/2/0,
    0/3/3, 1/3/0, 2/3/1, 3/3/4, 4/3/2,
    0/4/4, 1/4/2, 2/4/3, 3/4/0, 4/4/1
  }{
    \PutDigit{\c}{\r}{\txt}
  }

\end{tikzpicture}
    \caption{Orthogonal mate}
    \label{fig:intro_c}
\end{subfigure}
\hfill
\begin{subfigure}[t]{0.3\textwidth}
    \centering
\definecolor{cellred}{RGB}{229,115,115}
\definecolor{cellblue}{RGB}{100,181,246}
\definecolor{cellpurple}{RGB}{186,104,200}
\definecolor{cellyellow}{RGB}{255,241,118}
\definecolor{cellgreen}{RGB}{129,199,132}

\newcommand{\FillCell}[3]{%
  \pgfmathtruncatemacro{\yy}{4-#2}
  \ifcase#3\relax
    \path[fill=cellred,draw=none] (#1,\yy) rectangle ++(1,1);
  \or
    \path[fill=cellblue,draw=none] (#1,\yy) rectangle ++(1,1);
  \or
    \path[fill=cellpurple,draw=none] (#1,\yy) rectangle ++(1,1);
  \or
    \path[fill=cellyellow,draw=none] (#1,\yy) rectangle ++(1,1);
  \or
    \path[fill=cellgreen,draw=none] (#1,\yy) rectangle ++(1,1);
  \fi
}

\newcommand{\PutDigit}[3]{%
  \pgfmathtruncatemacro{\yy}{4-#2}
  \node[inner sep=0pt,font=\fontsize{18}{18}\selectfont] at ({#1+0.5},{\yy+0.5}) {#3};
}

\newcommand{\PutPair}[3]{%
  \pgfmathtruncatemacro{\yy}{4-#2}
  \node[inner sep=0pt,font=\fontsize{15}{15}\selectfont] at ({#1+0.5},{\yy+0.5}) {#3};
}

\newcommand{\DrawGrid}{%
  \foreach \k in {0,1,2,3,4,5} {
    \draw[black, line width=0.4pt] (\k,0) -- (\k,5);
    \draw[black, line width=0.4pt] (0,\k) -- (5,\k);
  }
}

\begin{tikzpicture}[x=0.92cm,y=0.92cm]
  \foreach \c/\r/\clr in {
    0/0/0, 1/0/1, 2/0/2, 3/0/3, 4/0/4,
    0/1/3, 1/1/4, 2/1/0, 3/1/1, 4/1/2,
    0/2/1, 1/2/2, 2/2/3, 3/2/4, 4/2/0,
    0/3/4, 1/3/0, 2/3/1, 3/3/2, 4/3/3,
    0/4/2, 1/4/3, 2/4/4, 3/4/0, 4/4/1
  }{
    \FillCell{\c}{\r}{\clr}
  }

  \DrawGrid

  \foreach \c/\r/\txt in {
    0/0/0\,0, 1/0/1\,1, 2/0/2\,4, 3/0/3\,2, 4/0/4\,3,
    0/1/1\,2, 1/1/2\,3, 2/1/3\,0, 3/1/4\,1, 4/1/0\,4,
    0/2/2\,1, 1/2/3\,4, 2/2/4\,2, 3/2/0\,3, 4/2/1\,0,
    0/3/3\,3, 1/3/4\,0, 2/3/0\,1, 3/3/1\,4, 4/3/2\,2,
    0/4/4\,4, 1/4/0\,2, 2/4/1\,3, 3/4/2\,0, 4/4/3\,1
  }{
    \PutPair{\c}{\r}{\txt}
  }

\end{tikzpicture}
    \caption{Overlay}
    \label{fig:intro_d}
\end{subfigure}

\caption{(a) A Latin square of order 5 decomposed into 5 non-overlapping transversals, with each
transversal highlighted in a separate colour.
(b) The orthogonal mate of the square in (a) produced by labelling the cells of each colour with a distinct symbol.
(c) The original square and orthogonal mate overlaid, showing that every pair of symbols appears exactly once.
}
\label{fig:intro}
\end{figure}

Say $L$ is the Latin square for which we want to find all orthogonal mates.
Instead of searching for orthogonal mates of $L$ cell-by-cell, Parker used an
approach that decomposes the problem into two stages.
The first stage finds all transversals of $L$, and the second
stage finds all ways of combining those transversals together in non-overlapping ways
to form a complete square.
Whenever $n$ non-overlapping transversals are found, an orthogonal
mate is constructed by Euler's observation (see Figure~\ref{fig:intro}), and consequently this approach is known as the
Euler--Parker algorithm for orthogonal mate construction.
Knuth estimates that the Euler--Parker approach 
of searching for orthogonal mates of a $10\times10$ square reduces the search cost
by an astonishing factor of $10^{12}$~\cite{Knuth2011TAOCP4A}.

Another method of searching for orthogonal pairs of Latin squares uses automated reasoning tools
such as satisfiability (SAT) solvers~\cite{kochemazov2016comparison}, constraint or integer programming solvers~\cite{Appa2006,Rubin2021}, or finite model searching tools~\cite{Ma2013}.
Such tools tend not to be competitive with the Euler--Parker approach if one only wants a pair of Latin squares and does not require them to be of any particular special form.
However, in many problems involving orthogonal Latin squares it is not enough to construct an arbitrary pair---one wants to find
a pair of orthogonal Latin squares with a special structure, and in such problems automated reasoning tools have proven to be extremely useful.
In this setting, an issue with the Euler--Parker approach is that it requires an initial
Latin square to start from.

For small orders, a significant proportion of Latin squares have orthogonal mates.  For example,
McKay et al.~\cite{McKay2006} estimate that around 61\% of Latin squares of order 10 have orthogonal mates.
In practice, one can find Latin squares with a mate
through random generation or by a theoretical construction.
However, when one seeks a pair of a special form, in order to use Euler--Parker one must find a square with a mate
but also with a mate \emph{of the special form}.  Such squares may be extremely rare and with
no obvious approach for finding them.  In such cases, automated reasoning tools like SAT solvers
can be effective when the extra constraints defining the special form can be provided as additional
constraints.  That way the solver is able to direct its search towards orthogonal Latin squares
of the specific form in question.

Despite the success of automated reasoning on many orthogonal Latin square problems (see Section~\ref{sec:related_work}), automated reasoning tools have no
knowledge of the Euler--Parker approach.  A workaround is to encode into the problem
instance that the symbols of one Latin square partition
the other Latin square into disjoint transversals (see Section~\ref{sec:encoding}).
This is essentially a ``declarative'' form of the Euler--Parker algorithm,
and in fact this encoding performs decently well in practice~\cite{BrightKeitaStevens2026Myrvold}.
The goal of our work is to show that it can be significantly beneficial to call an external
implementation of the Euler--Parker algorithm during solving,
rather than only relying on the declarative Euler--Parker orthogonality constraints.

In our approach, when the solver finds a partial solution that represents a complete Latin square~$L$, we call an implementation
of the Euler--Parker algorithm to determine if $L$ has an orthogonal mate.  If not, we tell the
solver to backtrack immediately.  In contrast, a pure SAT approach would
not distinguish between the square-finding and transversal-finding stages. When a Latin square~$L$ is found,
the solver will continue to try to find $n$ disjoint transversals of $L$,
but there is no guarantee this step will run to exhaustion:
if the solver encounters a conflict when looking for transversals, it will backtrack and in doing so
may choose to explore a new Latin square instead.
Even worse, in general not all transversals will be explored, because the solver will typically have transversals already
appearing in its partial solution.
If the transversals in the partial solution do not extend to $n$ disjoint transversals,
then an orthogonal mate of $L$ will not be found without backtracking,
even when $L$ has an orthogonal mate.
This drawback could be avoided by having the solver branch on variables encoding entries of $L$
before variables encoding transversals of $L$.
However, this still would not exploit the full Euler--Parker approach,
whose power comes from the fact that knowing all transversals of $L$
allows one to quickly construct a mate for $L$ or rule out its existence.

\subsection{Benchmarks}

As a case study for benchmarking our hybrid implementation combining SAT with a symbolic
Euler--Parker implementation, we evaluate the performance of our solver on a collection of
difficult orthogonal Latin square problems originating in the work of
Myrvold~\cite{Myrvold1999NegativeResults} in 1999 (see Section~\ref{sec:background_myrv}
for more details).
Myrvold was studying the notoriously difficult
problem of finding a collection of three mutually orthogonal Latin squares of order~10, also known as a $\MOLS{3}{10}$.
Despite attracting an immense amount of interest since at least the 1950s, the question of existence of a $\MOLS{3}{10}$
is still open and is considered to be an extremely difficult problem---%
Knuth lists it as a research problem of difficulty 50, his highest possible ranking~\cite[Ex.~15]{Knuth2011TAOCP4A}.

Myrvold considered the case of a $\MOLS{3}{10}$ in which one of the squares has a $4\times4$ Latin subsquare
and then considered the form that the other two squares (a $\MOLS{2}{10}$) must take on.
She showed that the $\MOLS{2}{10}$ must be one of twenty-eight possible forms,
and she was able to eliminate twenty of the twenty-eight possibilities by theoretical arguments.
However, the remaining eight cases proved to be more difficult and resisted any approach of ruling them out.
For over twenty-five years it was unknown if some more complicated argumentation could eliminate them,
or if $\MOLS{2}{10}$s existed in some or all of the eight cases.  This was the state of affairs until recently,
when Bright, Keita, and Stevens~\cite{BrightKeitaStevens2026Myrvold} used a SAT solver
to explicitly construct $\MOLS{2}{10}$s in all eight of Myrvold's remaining cases.

Bright, Keita, and Stevens~\cite{BrightKeitaStevens2026Myrvold} found examples in
Myrvold's eight cases using a pure SAT encoding
and the off-the-shelf SAT solver Kissat~\cite{biere2024kissat}.  Because SAT solvers
inherently have an element of randomness in their running times, for the purposes
of decreasing the amount of variance, each of the eight cases had
49 instances that were run independently, each with a different random seed.
They report that the SAT solver took a median of around 580,000 seconds to solve
the hardest of the eight cases, and 24 of the 49 instances in the hardest case
timed out after one week without a solution being found.

In contrast, in this paper we implement a new hybrid approach
that combines the SAT solver {\CaDiCaL}~\cite{cadical} with an external call to the
Euler--Parker algorithm (see Section~\ref{sec:prog_ep}).
The hybrid approach significantly improves the running time
of a pure SAT solver:
our hybrid version of {\CaDiCaL} solves the hardest of Myrvold's eight cases
in a median of about 5,100 seconds, over 100 times faster than Kissat.
Note that the SAT solver Kissat (used by Bright, Keita, and Stevens) is more effective than {\CaDiCaL}
at solving these instances.  Even still, our hybrid solver (based on
{\CaDiCaL}) dramatically outperformed both Kissat and pure {\CaDiCaL}, solving all the
instances in the hardest case in around 12 hours,
whereas Kissat only solved half of the hardest instances
within a week and pure {\CaDiCaL} solved none of the hardest instances within a week.
See Section~\ref{sec:results} for a summary of our results.

\section{Background} \label{sec:background}

In this section we give the background required for the rest of the paper.
In particular, we outline satisfiability solving in Section~\ref{sec:background_SAT},
linear Diophantine system solving as used by the Euler--Parker approach in Section~\ref{sec:background_EC},
and a description of Myrvold's $\MOLS{2}{10}$ instances that we use for benchmarking in Section~\ref{sec:background_myrv}.
We close the section with a description of related work using automated reasoning
tools on orthogonal Latin square problems in Section~\ref{sec:related_work}.

\subsection{Satisfiability solving} \label{sec:background_SAT}

The Boolean satisfiability problem (SAT) asks whether a Boolean formula admits a truth assignment that makes the formula evaluate to true.
It was the first problem shown to be NP-complete~\cite{Cook1971}, and it remains a central problem in computational complexity.
Perhaps surprisingly, SAT solving has also become a useful tool in computational discrete mathematics---the webpage
\href{https://sat4math.com}{sat4math.com} lists over 100 papers in which SAT solvers were used to solve problems in mathematics.
This success stems from the fact that over the last several decades
advances in SAT solving have produced highly effective general-purpose search tools~\cite{Ganesh2020}.
Their search capabilities have proved useful in a wide range of areas,
including discrete geometry~\cite{Heule2024,Subercaseaux2025,Subercaseaux2024},
finite geometry~\cite{LamPaper},
infinite graph theory~\cite{Subercaseaux2023},
and various kinds of puzzles~\cite{Bright2020}.

Modern SAT solvers typically require the input formula to be given in \emph{conjunctive normal form} (CNF), defined in terms of literals and clauses. A \emph{literal} is a Boolean variable~$p$ or its negation~$\lnot p$. A \emph{clause} is a disjunction of literals, and a formula is in \emph{CNF} if it is a conjunction of clauses. For example, $(p \lor q) \land p \land (\lnot p \lor q)$ is a CNF formula; its second clause is a \emph{unit clause} because it consists of one literal. We also use implication notation as shorthand, with $p \limp q$ denoting $\lnot p \lor q$. A \emph{satisfying assignment} is a truth assignment to the variables such that every clause evaluates to true.

To apply SAT solving to a search problem, one encodes the problem as a CNF formula so that the original problem has a solution if and only if the formula is satisfiable. A satisfying assignment can then be decoded into a solution of the original search problem. Conversely, if the formula is unsatisfiable, then the original search problem has no solution.

\subsubsection{Programmatic SAT with IPASIR-UP} \label{sec:background_IPASIR}

{\CaDiCaL} supports a programmatic interface based on IPASIR-UP~\cite{ipasirup}, which allows user-defined callback functions to interact with the solver during search. In particular, the interface makes it possible to monitor assignments to selected variables, detect backtracking events, and add clauses dynamically based on information not encoded directly in CNF\@.

In our setting, the IPASIR-UP interface is used to detect when all variables encoding a Latin square have been assigned.
At that point, we invoke an Euler--Parker implementation to determine whether the completed square admits an orthogonal mate. If the square does not have an orthogonal mate, the solver is then given a clause that blocks the square from being a solution and forces the solver to backtrack.

The principal IPASIR-UP callbacks used in our method are
\begin{itemize}[nolistsep,noitemsep]
    \item \texttt{notify\_assignment()}, which reports when a variable has been assigned;
    \item \texttt{notify\_backtrack()}, which reports when a backtrack occurs, causing some assigned variables to become unassigned; and
    \item \texttt{has\_external\_clause()}, which indicates that an externally generated clause is ready to be added to the solver.  In our method,
    whenever the Euler--Parker algorithm fails to construct a mate of a Latin square $L$, we add a clause blocking $L$ from being considered again.
\end{itemize}

\subsubsection{Combining SAT solving and symbolic computation} \label{sec:sc2}

Interfaces like IPASIR-UP are useful when part of the instance is more naturally handled outside the base SAT encoding.
For example, one may invoke a computer algebra system (CAS) on a structured subproblem determined by the current partial assignment.
This ``SAT+CAS'' paradigm has shown to be beneficial in a number of mathematical problems~\cite{Bright2022}.

More generally, combinations of satisfiability checking (SC) and symbolic computation (SC) have
led to a number of successes, resulting in what has become known as the ``SC$^2$'' paradigm~\cite{England2022a}.
The paradigm had its inception around 2015, when \'Abrah\'am~\cite{Abraham2015} and Zulkoski et al.~\cite{Zulkoski2015}
proposed combining the fields of satisfiability checking and symbolic computation to solve mathematical problems.
Both fields were mature with well-established methods for solving mathematical problems, but there had been limited
interaction between the fields up to that point, despite the fields' complementary strengths.
Satisfiability checking is effective at general search and learning, while symbolic computation is
effective at solving problems that admit clever solutions by exploiting mathematical structure.
The SC$^2$ approach therefore excels at solving problems that can benefit from both effective search
and exploiting mathematical structure.

The SC$^2$ workshop, now in its eleventh year, hosts work that combines satisfiability checking
and symbolic computation.  What is particularly notable is the sheer variety of problems that
benefit from the SC$^2$ framework.  For example, SC$^2$ approaches have been used on
graph enumeration up to isomorphism~\cite{Li2022,ijcai2024}, factorizing integers with known bits of the factors~\cite{Ajani2023,Ajani2024},
searching for hash function collisions~\cite{Alamgir2024}, and proving theorems in extremal combinatorics~\cite{Ahmed2025}.

\subsection{Linear Diophantine system solving} \label{sec:background_EC}

For a Latin square $L$ of order $n$, an orthogonal mate exists if and only if the cells of $L$ can be partitioned into $n$ disjoint transversals.
This leads naturally to an exact cover viewpoint: one seeks a collection of transversals covering each cell exactly once.
In general, an \emph{exact cover} of a set $X$ is a family of sets $S_1$, $\dotsc$, $S_N$ where $\bigcup_{i=1}^N S_i = X$, and the sets $S_i$ are pairwise disjoint ($S_i\cap S_j=\emptyset$ for all $1\leq i<j\leq N$).  The corresponding decision problem is:
given~$X$ and a family of sets $S_1$, $\dotsc$, $S_N$, does there exist a subfamily of the family of sets forming an exact cover of $X$?

The Euler--Parker algorithm uses a two-stage approach in order to construct the orthogonal mates of a Latin square $L$ of order~$n$.  First, all transversals of $L$ are found.  Each transversal is represented as a set of $n$ cells.  Second, $n$ pairwise disjoint transversals of $L$ are found.  These $n$ transversals form an exact cover of the set of all $n^2$ cells $\{(0,0),(0,1),\dotsc,(n-1,n-1)\}$.

Our implementation of the Euler--Parker algorithm uses the library {\libexact}~\cite{kaski2008libexact} in order to solve
the exact cover problems.
{\libexact} is a library that finds all solutions of the linear Diophantine system
\begin{equation*}
A \vec{x} = \vec{b}, \qquad \mathbf{0} \leq \vec{x} \leq \vec{u}
\end{equation*}
where $A = [a_{ij}]$ is a specified $M \times N$ matrix with entries in $\{0,1\}$, $\vec{b}$ and $\vec{u}$ are specified column vectors of positive integers, and the inequalities hold entrywise.
A solution of the system is represented by the column vector $\vec{x} = (x_1;\dots;x_N)$ of nonnegative integers.
For the purposes of solving the
exact cover problem described above,
we take $\vec{b}$ and $\vec{u}$ to be $\mathbf{1}$, the vector of all 1s of appropriate dimension.

Combinatorially, each row of $A$ represents a constraint (e.g., that a specific cell must appear in a cover)
and each column of $A$ represents a choice (e.g., if a transversal of $L$ will be used in a cover or not).
The column $j$ covers row~$i$ precisely when $a_{ij}=1$.
The equation $A\vec{x}=\mathbf{1}$ requires that each row~$i$ be covered exactly once,
while the bounds $0\leq x_j\leq 1$ require that each column $j$ be used at most once.
Each variable~$x_j$ records if the corresponding choice is to be used in a cover,
with $x_j$ corresponding to $S_j$, where $S_1$, $\dotsc$, $S_N$ is the family of all possible
sets that could appear in a cover.  The set~$S_j$ will be used in the cover when~$x_j=1$, and the set
$S_j$ will not be used in the cover when~$x_j=0$.

\subsection{Myrvold's $\MOLS{2}{10}$ results} \label{sec:background_myrv}

In 1999, Myrvold studied the problem of extending an orthogonal pair of $10 \times 10$ Latin squares to an orthogonal triple under the assumption that one of the squares contains a $4 \times 4$ Latin subsquare~\cite{Myrvold1999NegativeResults}. She derived a highly restrictive classification of the possible orthogonal pair types in this setting, ruling out twenty of twenty-eight cases. The remaining eight cases were left unresolved. Subsequent work by Bright et al.~\cite{BrightKeitaStevens2026Myrvold} showed that orthogonal pairs do exist in each of these eight cases, although the existence of an orthogonal triple remains unknown. Consequently, any argument excluding the remaining cases must necessarily involve the third square, rather than only the orthogonal pair.

The full details of Myrvold's framework can be found in~\cite{BrightKeitaStevens2026Myrvold,Myrvold1999NegativeResults}, and
we review only the aspects needed here.
Instead of working directly with a $\MOLS{2}{10}$, Myrvold considers a \emph{transversal representation pair} (TRP) of Latin squares $(P,Q)$.
In such a pair, each row of $Q$ represents a transversal of $P$, and the $n$ rows of $Q$
decompose the cells
of $P$ into $n$ disjoint transversals.
Thus $P$ and $Q$ are not themselves orthogonal mates, but they imply the existence of an orthogonal pair and
it is straightforward to construct a $\MOLS{2}{10}$ from $(P,Q)$ and vice versa.

Myrvold assumes that there exists a Latin square $L$ whose transversals are represented by both $P$ and $Q$. The square $L$ is taken to contain a $4 \times 4$ Latin subsquare $\Omega$ in its lower-right corner, using the symbols $\{0,1,2,3\}$. This assumption induces a colouring on the entries of $P$ and $Q$. Entries containing symbols in $\{0,1,2,3\}$ are called \emph{white}. Among the entries in the first six columns, exactly two nonwhite entries from each column are coloured \emph{dark}, and the remaining nonwhite entries are called \emph{light}.

There are four transversal types, denoted $p_1$, $\dotsc$, $p_4$, classified by the number of white entries in the last four columns. A transversal of type $p_i$ contains exactly $i$ white entries in its last four columns and exactly $2i-2$ dark entries.
Myrvold studied the possible ways in which $L$ could be decomposed into 10 transversals and showed that there were
exactly seven general types (determined by the number of transversals of each type $p_1$, $\dotsc$, $p_4$) which she denoted R, S, T, U, V, W, and X\@.
For example, a decomposition of $L$ using eight transversals of type $p_1$ and two transversals of type $p_4$ was said to be of type R\@.

Under the assumption that $L$ belongs to an orthogonal triple and therefore is decomposable into 10 disjoint transversals
in two ways, Myrvold showed that the only pairs of potentially feasible decomposition types for $L$
were UU, SX, UW, WW, VX, UX, WX, and XX\@.  She was able to theoretically rule out each of the other twenty
possible pairings (e.g., $L$ cannot be decomposed into R and S simultaneously),
although she was unable to confirm or rule out any of the pairings UU, SX, UW, WW, VX, UX, WX, and XX\@.
It turns out this is because transversal representation pairs in these eight cases actually exist.
An example of a type-UW transversal representation pair is shown in Figure~\ref{fig:trp}.
Note that $(P,Q)$ does \emph{not} extend to a triple $(P,Q,L)$,
and it appears to be a very difficult problem to find such a triple.
The most we can say is that
there exists a $4\times4$ Latin subsquare $\Omega$ of~$L$
compatible with $(P,Q)$,
because the SAT encoding used to generate the pair
included ``$\Omega$ compatibility'' constraints (see~\cite[Sec.~4.4]{BrightKeitaStevens2026Myrvold}).

\begin{figure}[htbp]
\centering
\resizebox{0.75\textwidth}{!}{%
\begin{tikzpicture}[x=0.62cm,y=0.62cm]
  \definecolor{lightcell}{HTML}{98C5FF}
  \definecolor{darkcell}{HTML}{1D84C4}

  \newcommand{\mcell}[5]{%
    \fill[#4] (#1,#2) rectangle ++(1,1);
    \node[text=#5,font=\rmfamily\fontsize{14}{14}\selectfont] at (#1+0.5,#2+0.45) {#3};
  }

  \node[font=\rmfamily\fontsize{18}{18}\selectfont] at (5,10.55) {type U};
  \mcell{0}{9}{0}{white}{black}
  \mcell{1}{9}{2}{white}{black}
  \mcell{2}{9}{3}{white}{black}
  \mcell{3}{9}{4}{lightcell}{black}
  \mcell{4}{9}{5}{lightcell}{black}
  \mcell{5}{9}{6}{lightcell}{black}
  \mcell{6}{9}{1}{white}{black}
  \mcell{7}{9}{7}{lightcell}{black}
  \mcell{8}{9}{8}{lightcell}{black}
  \mcell{9}{9}{9}{lightcell}{black}
  \mcell{0}{8}{1}{white}{black}
  \mcell{1}{8}{0}{white}{black}
  \mcell{2}{8}{4}{lightcell}{black}
  \mcell{3}{8}{9}{lightcell}{black}
  \mcell{4}{8}{7}{lightcell}{black}
  \mcell{5}{8}{2}{white}{black}
  \mcell{6}{8}{6}{lightcell}{black}
  \mcell{7}{8}{5}{lightcell}{black}
  \mcell{8}{8}{3}{white}{black}
  \mcell{9}{8}{8}{lightcell}{black}
  \mcell{0}{7}{3}{white}{black}
  \mcell{1}{7}{9}{lightcell}{black}
  \mcell{2}{7}{8}{lightcell}{black}
  \mcell{3}{7}{2}{white}{black}
  \mcell{4}{7}{4}{lightcell}{black}
  \mcell{5}{7}{1}{white}{black}
  \mcell{6}{7}{0}{white}{black}
  \mcell{7}{7}{6}{lightcell}{black}
  \mcell{8}{7}{7}{lightcell}{black}
  \mcell{9}{7}{5}{lightcell}{black}
  \mcell{0}{6}{4}{lightcell}{black}
  \mcell{1}{6}{1}{white}{black}
  \mcell{2}{6}{7}{lightcell}{black}
  \mcell{3}{6}{0}{white}{black}
  \mcell{4}{6}{3}{white}{black}
  \mcell{5}{6}{5}{lightcell}{black}
  \mcell{6}{6}{9}{lightcell}{black}
  \mcell{7}{6}{8}{lightcell}{black}
  \mcell{8}{6}{6}{lightcell}{black}
  \mcell{9}{6}{2}{white}{black}
  \mcell{0}{5}{5}{lightcell}{black}
  \mcell{1}{5}{8}{lightcell}{black}
  \mcell{2}{5}{0}{white}{black}
  \mcell{3}{5}{1}{white}{black}
  \mcell{4}{5}{2}{white}{black}
  \mcell{5}{5}{7}{lightcell}{black}
  \mcell{6}{5}{3}{white}{black}
  \mcell{7}{5}{9}{lightcell}{black}
  \mcell{8}{5}{4}{lightcell}{black}
  \mcell{9}{5}{6}{lightcell}{black}
  \mcell{0}{4}{8}{lightcell}{black}
  \mcell{1}{4}{3}{white}{black}
  \mcell{2}{4}{1}{white}{black}
  \mcell{3}{4}{6}{lightcell}{black}
  \mcell{4}{4}{9}{lightcell}{black}
  \mcell{5}{4}{0}{white}{black}
  \mcell{6}{4}{4}{lightcell}{black}
  \mcell{7}{4}{2}{white}{black}
  \mcell{8}{4}{5}{lightcell}{black}
  \mcell{9}{4}{7}{lightcell}{black}
  \mcell{0}{3}{2}{white}{black}
  \mcell{1}{3}{7}{darkcell}{white}
  \mcell{2}{3}{6}{lightcell}{black}
  \mcell{3}{3}{8}{darkcell}{white}
  \mcell{4}{3}{0}{white}{black}
  \mcell{5}{3}{9}{lightcell}{black}
  \mcell{6}{3}{5}{lightcell}{black}
  \mcell{7}{3}{3}{white}{black}
  \mcell{8}{3}{1}{white}{black}
  \mcell{9}{3}{4}{lightcell}{black}
  \mcell{0}{2}{9}{lightcell}{black}
  \mcell{1}{2}{6}{lightcell}{black}
  \mcell{2}{2}{2}{white}{black}
  \mcell{3}{2}{5}{darkcell}{white}
  \mcell{4}{2}{1}{white}{black}
  \mcell{5}{2}{8}{darkcell}{white}
  \mcell{6}{2}{7}{lightcell}{black}
  \mcell{7}{2}{4}{lightcell}{black}
  \mcell{8}{2}{0}{white}{black}
  \mcell{9}{2}{3}{white}{black}
  \mcell{0}{1}{6}{darkcell}{white}
  \mcell{1}{1}{4}{darkcell}{white}
  \mcell{2}{1}{5}{darkcell}{white}
  \mcell{3}{1}{7}{lightcell}{black}
  \mcell{4}{1}{8}{darkcell}{white}
  \mcell{5}{1}{3}{white}{black}
  \mcell{6}{1}{2}{white}{black}
  \mcell{7}{1}{1}{white}{black}
  \mcell{8}{1}{9}{lightcell}{black}
  \mcell{9}{1}{0}{white}{black}
  \mcell{0}{0}{7}{darkcell}{white}
  \mcell{1}{0}{5}{lightcell}{black}
  \mcell{2}{0}{9}{darkcell}{white}
  \mcell{3}{0}{3}{white}{black}
  \mcell{4}{0}{6}{darkcell}{white}
  \mcell{5}{0}{4}{darkcell}{white}
  \mcell{6}{0}{8}{lightcell}{black}
  \mcell{7}{0}{0}{white}{black}
  \mcell{8}{0}{2}{white}{black}
  \mcell{9}{0}{1}{white}{black}
  \draw[line width=0.55pt] (0,0) -- (0,10);
  \draw[line width=0.55pt] (0, 0) -- (10, 0);
  \draw[line width=0.55pt] (1,0) -- (1,10);
  \draw[line width=0.55pt] (0, 1) -- (10, 1);
  \draw[line width=0.55pt] (2,0) -- (2,10);
  \draw[line width=0.55pt] (0, 2) -- (10, 2);
  \draw[line width=0.55pt] (3,0) -- (3,10);
  \draw[line width=0.55pt] (0, 3) -- (10, 3);
  \draw[line width=0.55pt] (4,0) -- (4,10);
  \draw[line width=0.55pt] (0, 4) -- (10, 4);
  \draw[line width=0.55pt] (5,0) -- (5,10);
  \draw[line width=0.55pt] (0, 5) -- (10, 5);
  \draw[line width=0.55pt] (6,0) -- (6,10);
  \draw[line width=0.55pt] (0, 6) -- (10, 6);
  \draw[line width=0.55pt] (7,0) -- (7,10);
  \draw[line width=0.55pt] (0, 7) -- (10, 7);
  \draw[line width=0.55pt] (8,0) -- (8,10);
  \draw[line width=0.55pt] (0, 8) -- (10, 8);
  \draw[line width=0.55pt] (9,0) -- (9,10);
  \draw[line width=0.55pt] (0, 9) -- (10, 9);
  \draw[line width=0.55pt] (10,0) -- (10,10);
  \draw[line width=0.55pt] (0, 10) -- (10, 10);

  \node[font=\rmfamily\fontsize{18}{18}\selectfont] at (17,10.55) {type W};
  \mcell{12}{9}{0}{white}{black}
  \mcell{13}{9}{1}{white}{black}
  \mcell{14}{9}{8}{lightcell}{black}
  \mcell{15}{9}{3}{white}{black}
  \mcell{16}{9}{7}{lightcell}{black}
  \mcell{17}{9}{9}{lightcell}{black}
  \mcell{18}{9}{2}{white}{black}
  \mcell{19}{9}{4}{lightcell}{black}
  \mcell{20}{9}{5}{lightcell}{black}
  \mcell{21}{9}{6}{lightcell}{black}
  \mcell{12}{8}{1}{white}{black}
  \mcell{13}{8}{6}{lightcell}{black}
  \mcell{14}{8}{7}{lightcell}{black}
  \mcell{15}{8}{4}{lightcell}{black}
  \mcell{16}{8}{2}{white}{black}
  \mcell{17}{8}{0}{white}{black}
  \mcell{18}{8}{8}{lightcell}{black}
  \mcell{19}{8}{3}{white}{black}
  \mcell{20}{8}{9}{lightcell}{black}
  \mcell{21}{8}{5}{lightcell}{black}
  \mcell{12}{7}{2}{white}{black}
  \mcell{13}{7}{5}{lightcell}{black}
  \mcell{14}{7}{4}{lightcell}{black}
  \mcell{15}{7}{6}{lightcell}{black}
  \mcell{16}{7}{3}{white}{black}
  \mcell{17}{7}{1}{white}{black}
  \mcell{18}{7}{7}{lightcell}{black}
  \mcell{19}{7}{9}{lightcell}{black}
  \mcell{20}{7}{8}{lightcell}{black}
  \mcell{21}{7}{0}{white}{black}
  \mcell{12}{6}{3}{white}{black}
  \mcell{13}{6}{0}{white}{black}
  \mcell{14}{6}{2}{white}{black}
  \mcell{15}{6}{7}{lightcell}{black}
  \mcell{16}{6}{9}{lightcell}{black}
  \mcell{17}{6}{6}{lightcell}{black}
  \mcell{18}{6}{5}{lightcell}{black}
  \mcell{19}{6}{8}{lightcell}{black}
  \mcell{20}{6}{4}{lightcell}{black}
  \mcell{21}{6}{1}{white}{black}
  \mcell{12}{5}{5}{lightcell}{black}
  \mcell{13}{5}{9}{lightcell}{black}
  \mcell{14}{5}{6}{lightcell}{black}
  \mcell{15}{5}{0}{white}{black}
  \mcell{16}{5}{1}{white}{black}
  \mcell{17}{5}{3}{white}{black}
  \mcell{18}{5}{4}{lightcell}{black}
  \mcell{19}{5}{7}{lightcell}{black}
  \mcell{20}{5}{2}{white}{black}
  \mcell{21}{5}{8}{lightcell}{black}
  \mcell{12}{4}{6}{darkcell}{white}
  \mcell{13}{4}{8}{lightcell}{black}
  \mcell{14}{4}{3}{white}{black}
  \mcell{15}{4}{5}{darkcell}{white}
  \mcell{16}{4}{4}{lightcell}{black}
  \mcell{17}{4}{2}{white}{black}
  \mcell{18}{4}{9}{lightcell}{black}
  \mcell{19}{4}{0}{white}{black}
  \mcell{20}{4}{1}{white}{black}
  \mcell{21}{4}{7}{lightcell}{black}
  \mcell{12}{3}{7}{darkcell}{white}
  \mcell{13}{3}{2}{white}{black}
  \mcell{14}{3}{1}{white}{black}
  \mcell{15}{3}{9}{lightcell}{black}
  \mcell{16}{3}{8}{darkcell}{white}
  \mcell{17}{3}{5}{lightcell}{black}
  \mcell{18}{3}{3}{white}{black}
  \mcell{19}{3}{6}{lightcell}{black}
  \mcell{20}{3}{0}{white}{black}
  \mcell{21}{3}{4}{lightcell}{black}
  \mcell{12}{2}{8}{lightcell}{black}
  \mcell{13}{2}{4}{darkcell}{white}
  \mcell{14}{2}{9}{darkcell}{white}
  \mcell{15}{2}{2}{white}{black}
  \mcell{16}{2}{0}{white}{black}
  \mcell{17}{2}{7}{lightcell}{black}
  \mcell{18}{2}{1}{white}{black}
  \mcell{19}{2}{5}{lightcell}{black}
  \mcell{20}{2}{6}{lightcell}{black}
  \mcell{21}{2}{3}{white}{black}
  \mcell{12}{1}{9}{lightcell}{black}
  \mcell{13}{1}{3}{white}{black}
  \mcell{14}{1}{0}{white}{black}
  \mcell{15}{1}{8}{darkcell}{white}
  \mcell{16}{1}{5}{lightcell}{black}
  \mcell{17}{1}{4}{darkcell}{white}
  \mcell{18}{1}{6}{lightcell}{black}
  \mcell{19}{1}{1}{white}{black}
  \mcell{20}{1}{7}{lightcell}{black}
  \mcell{21}{1}{2}{white}{black}
  \mcell{12}{0}{4}{lightcell}{black}
  \mcell{13}{0}{7}{darkcell}{white}
  \mcell{14}{0}{5}{darkcell}{white}
  \mcell{15}{0}{1}{white}{black}
  \mcell{16}{0}{6}{darkcell}{white}
  \mcell{17}{0}{8}{darkcell}{white}
  \mcell{18}{0}{0}{white}{black}
  \mcell{19}{0}{2}{white}{black}
  \mcell{20}{0}{3}{white}{black}
  \mcell{21}{0}{9}{lightcell}{black}
  \draw[line width=0.55pt] (12,0) -- (12,10);
  \draw[line width=0.55pt] (12, 0) -- (22, 0);
  \draw[line width=0.55pt] (13,0) -- (13,10);
  \draw[line width=0.55pt] (12, 1) -- (22, 1);
  \draw[line width=0.55pt] (14,0) -- (14,10);
  \draw[line width=0.55pt] (12, 2) -- (22, 2);
  \draw[line width=0.55pt] (15,0) -- (15,10);
  \draw[line width=0.55pt] (12, 3) -- (22, 3);
  \draw[line width=0.55pt] (16,0) -- (16,10);
  \draw[line width=0.55pt] (12, 4) -- (22, 4);
  \draw[line width=0.55pt] (17,0) -- (17,10);
  \draw[line width=0.55pt] (12, 5) -- (22, 5);
  \draw[line width=0.55pt] (18,0) -- (18,10);
  \draw[line width=0.55pt] (12, 6) -- (22, 6);
  \draw[line width=0.55pt] (19,0) -- (19,10);
  \draw[line width=0.55pt] (12, 7) -- (22, 7);
  \draw[line width=0.55pt] (20,0) -- (20,10);
  \draw[line width=0.55pt] (12, 8) -- (22, 8);
  \draw[line width=0.55pt] (21,0) -- (21,10);
  \draw[line width=0.55pt] (12, 9) -- (22, 9);
  \draw[line width=0.55pt] (22,0) -- (22,10);
  \draw[line width=0.55pt] (12, 10) -- (22, 10);
  
\draw[red, line width=1.2pt] (0.03,9.03) rectangle (9.97,9.97);

\foreach \x/\y in {
  12/9,
  13/3,
  14/4,
  15/8,
  16/1,
  17/6,
  18/2,
  19/5,
  20/7,
  21/0
}{
  \draw[red, line width=1.2pt] (\x+0.03,\y+0.03) rectangle (\x+0.97,\y+0.97);
}
  
\end{tikzpicture}
}
\caption{A type-UW transversal representation pair, with white, light, and dark colouring.
Each row of each square represents a transversal of the other square.  For example,
consider the first row of the first square $[0,2,3,4,\dotsc]$.  Going through the second
square column-by-column and extracting the cell from each column containing these
symbols in sequence yields the cells $(0,0)$, $(6,1)$, $(5,2)$, $(1,3)$, $\dotsc$
which form a transversal of the second square
(the cells highlighted with a red border).
}
\label{fig:trp}
\end{figure}

\subsection{Related work} \label{sec:related_work}

Orthogonal Latin square problems have 
been approached with several optimization and search methods. Appa, Mourtos, and Magos~\cite{Appa2002} investigated the orthogonal Latin squares problem using a combination of integer and constraint programming. Appa, Magos, and Mourtos~\cite{Appa2004a,Appa2004b} later used integer programming both to search for $\MOLS{2}{n}$ and to prove the nonexistence of $\MOLS{2}{6}$. In later work, they studied $\MOLS{2}{n}$ and $\MOLS{3}{n}$ more generally~\cite{Appa2006}. Rubin, Bright, Cheung, and Stevens~\cite{Rubin2021} compared integer and constraint programming encodings for $\MOLS{2}{n}$, and found that an indexing-based constraint programming formulation performed well on orders $5 \le n \le 12$ and $\MOLS{3}{n}$ for $n \leq 9$. Ma and Zhang~\cite{Ma2013} applied finite model generation to the search for $\MOLS{2}{n}$.

SAT-based methods have also been used extensively for orthogonal Latin square variants. Zaikin, Kochemazov, and Semenov~\cite{Kochemazov2016} compared several SAT encodings for systems of orthogonal Latin squares and diagonal orthogonal Latin squares. Zaikin and Kochemazov~\cite{Zaikin2015} (and later in conjunction with Zhuravlev and Vatutin~\cite{Zaikin2016b}) studied diagonal Latin square systems and searched for pairs and triples of mutually orthogonal diagonal Latin squares of order 10 using SAT-based methods and volunteer computing. Vatutin, Zaikin, Manzyuk, and Nikitina~\cite{Vatutin2021} later applied a cube-and-conquer approach to the search for orthogonal Latin squares.

Lu, Liu, and Zhang~\cite{Lu2011} searched for doubly self-orthogonal Latin squares using SAT, constraint programming, and also developed a custom exhaustive search algorithm based on exact cover.
Falc\'on, Falc\'on, and N\'u\~nez~\cite{Falcon2017} computed totally symmetric and totally conjugate orthogonal partial Latin squares using a SAT solver.
Huang, Liu, Ge, Ma, and Zhang~\cite{Huang2019} investigated orthogonal golf designs via satisfiability testing.
Jin, Lv, Ge, Ma, and Zhang~\cite{Jin2021} used SAT to study Costas Latin squares and orthogonal Costas Latin squares.
More broadly, Zhang~\cite{Zhang1997} surveys Latin square problems as a natural application area for SAT encodings.

Zaikin, Vatutin, and Bright used a SAT solver with the same orthogonality encoding that we use in this paper to enumerate all
extended self-orthogonal diagonal Latin squares of orders up to 10~\cite{zaikin2025enumerating}.
Bright, Keita, and Stevens also used this orthogonality encoding to re-enumerate all orthogonal pairs of Latin squares of order $10$ whose associated nets have at least two nontrivial relations~\cite{bright2025}.

\section{A Hybrid SAT + Euler--Parker Approach} \label{sec:implementation}

In this section, we describe our hybrid SAT + Euler--Parker approach.
First, we describe the basic SAT encoding in Section~\ref{sec:encoding}.
Next, we describe how we implement the Euler--Parker algorithm
using the {\libexact} library in Section~\ref{sec:prog_ep},
and also describe how the Euler--Parker algorithm is programmatically called
while the SAT solver is solving.
Lastly, we describe how to specialize our hybrid approach specifically
to the $\MOLS{2}{10}$ cases considered by Myrvold in Section~\ref{sec:hybrid_myrvold}.

\subsection{SAT encoding} \label{sec:encoding}

First, we describe the SAT encoding used to ensure
an $n\times n$ matrix $P$ is a Latin square.  To represent
the entries of $P$, we introduce variables $P_{i,j,k}$ for $0\leq i,j,k<n$
denoting that cell $(i,j)$ of $P$ contains symbol $k$.
To ensure that $P$ is a Latin square, we impose \emph{exactly-one} constraints requiring that each cell contains exactly one symbol and that each symbol appears exactly once in each row and column.
Exactly-one constraints on $n$ variables can be straightforwardly encoded into CNF using a single \emph{at-least-one} clause of length~$n$ and
$\binom{n}{2}$ \emph{at-most-one} clauses of length 2,
but Bright et al.~\cite{BrightKeitaStevens2026Myrvold} found that using an exactly-one
CNF encoding based on the ``totalizer'' encoding~\cite{totalizer} performed better in practice, 
and so this is the encoding we use in our work.
We denote a cardinality constraint saying that exactly $k$ of the variables
in a set $S$ are true by $\sum_{s\in S}s=k$.  Then for example enforcing that cell $(i,j)$
of $P$ has exactly one symbol can then be done with the exactly-one constraint $\sum_{0 \le k < n} P_{i,j,k}=1$.

Orthogonality of two squares $P$ and $R$ can be encoded directly
into the SAT instance, but a naive encoding
uses $\Theta(n^6)$ clauses of length 4~\cite[Eq.~4]{bright2025}.
In practice, a better way of enforcing the orthogonality of $P$ and $R$
is to instead assert that for all symbols $0\leq k<n$
the cells containing the symbol~$k$ in~$R$ form a transversal in $P$.  In order to do this,
we introduce
a new square $Q$ whose $k$th row represents the symbols of $P$
contained in the cells having symbol $k$ in $R$.  This is done with the constraints
\[ (R_{i,j,k}\land P_{i,j,l})\rightarrow Q_{k,j,l} \qquad\text{for all $0\leq i,j,k,l<n$.} \]
Once $Q$ has been introduced, asserting that the cells with symbol $k$ in $R$
form a transversal of~$P$ is done by asserting that each row of $Q$ contains each symbol exactly once.
The transversals formed in this way are necessarily
disjoint, so the existence of~$Q$ ultimately implies
that $R$ is an orthogonal mate of~$P$.  In fact, $Q$ itself will be a Latin square,
and constraints enforcing this are additionally included in the encoding.
Bright et~al.~\cite{BrightKeitaStevens2026Myrvold} derive additional structure on the squares
in terms of a composition operation.
They prove that when $P$, $Q$, and~$R$ are Latin squares the composition
equation $QR=P$ is equivalent to $(P,R)$ being an orthogonal pair
and $(P,Q)$ being a transversal representation pair.

\subsection{Programmatic Euler--Parker} \label{sec:prog_ep}

The orthogonality encoding described in Section~\ref{sec:encoding}
can be considered a ``static'' implementation of the Euler--Parker approach.
However, a standard backtracking SAT solver
will never stop to compute the set of \emph{all} transversals of a square and therefore (as described in Section~\ref{sec:intro}) will not
fully exploit the Euler--Parker approach.  This motivated
us to augment a SAT solver with a programmatic implementation of the Euler--Parker
algorithm that can be called whenever the SAT solver finds a complete Latin square
during the search.

The hybrid method still uses the SAT encoding from Section~\ref{sec:encoding}
in order to direct the solver towards orthogonal squares,
but it also has an external Euler--Parker subroutine added in conjunction with the CNF encoding.
Whenever the SAT solver reaches a complete Latin square, the programmatic Euler--Parker 
implementation is invoked to determine whether that square admits an orthogonal mate.
If the square is determined to not have an orthogonal mate by Euler--Parker,
the solver will immediately backtrack (which is earlier than it would have otherwise)
and otherwise an orthogonal mate is produced.
We now describe how we implemented the two stages of the Euler--Parker method using
two calls to the library {\libexact}.
In the following, say $L$ is a Latin square of order $n$ and we want
to compute all orthogonal mates of $L$.

\subsubsection{Stage 1: Construct all transversals} \label{sec:exact_cover_s1}

A transversal of $L$ is a set of $n$ cells containing exactly one cell from each row, exactly one cell from each column, and
with $L$ containing each symbol exactly once amongst the transversal's cells.
Constructing a transversal naturally reduces to solving a linear Diophantine system
where a solution of the system provides a transversal $t$ of $L$.
Let $x_{i,j}\in\{0,1\}$ be a variable that will be 1 when cell $(i,j)$ appears in $t$.
Since $t$ contains exactly one cell from row~$i$, we have $\sum_{j=0}^{n-1}x_{i,j}=1$; since
$t$ contains exactly one cell from column~$j$, we have $\sum_{i=0}^{n-1}x_{i,j}=1$; and since $t$ contains
an occurrence of symbol~$k$ in $L$ exactly once, we have $\sum_{L[i,j]=k}x_{i,j}=1$.

Since there are $n$ rows, $n$ columns, and $n$ symbols, this provides a
linear Diophantine system with $n^2$ variables and $3n$ equations.  Given $L$,
our implementation symbolically forms the above system and uses {\libexact} to find all
integer 0--1 solutions of the system.  Each solution is transformed into a transversal
and the input to the second stage is the set of all transversals of $L$.

\subsubsection{Stage 2: Construct $n$ disjoint transversals} \label{sec:exact_cover_s2}

Given a set $T$ of transversals of a Latin square $L$, constructing a set of $n$ disjoint transversals
also naturally reduces to solving a linear Diophantine system, where a solution of the system provides
$n$ disjoint transversals $T^*\subseteq T$.
Let $x_t\in\{0,1\}$ be a variable that will be 1 when transversal $t\in T$
is one of the $n$ transversals in $T^*$.
Since every cell $(i,j)$ appears in exactly one transversal in $T^*$
(because the transversals are disjoint and there are $n$ of them), we have the equations
\[ \sum_{t\in T,(i,j)\in t}x_t=1 \qquad \text{for every cell $(i,j)$}. \]
This is a linear system with $\lvert T\rvert$ variables and $n^2$ equations. For $n=10$, we found that $\lvert T\rvert$ is approximately $825$ on average, but it grows rapidly with $n$, reaching over 3 million for $n=15$. 
Our implementation symbolically forms this system and uses {\libexact} to solve it
exhaustively.  Each solution has exactly $n$ variables with $x_t=1$, providing $n$ disjoint transversals.
Denoting these $n$
transversals as $t_0$, $\dotsc$, $t_{n-1}$, we create an orthogonal mate to $L$ by
assigning symbol $i$ to the cells of $t_i$ for all $0\leq i<n$.

\subsubsection{Injecting Euler--Parker into the SAT solver} \label{sec:hybrid_ipasir}

The hybrid aspect of the method comes from using IPASIR-UP to call {\libexact} to
perform the above two-stage process during SAT search.
Using the callback routines described in Section~\ref{sec:background_IPASIR}, the interface tracks the assignment status of the variables encoding a Latin square.
Once all variables encoding a square have been assigned, the interface takes the complete square (say $L$) and begins stage 1.
If fewer than $n$ transversals are found, then $L$ cannot admit an orthogonal mate.
In this case, the interface constructs a blocking clause from the literals encoding $L$ and returns it to the SAT solver.
This forces the solver to backtrack and prevents $L$ from being considered again.

If at least $n$ transversals are found, they are passed to stage 2, which attempts to select a disjoint set of $n$ transversals.
If this second stage fails, the complete square $L$ is again blocked.
If it succeeds, then an orthogonal mate has been found.
If there happen to be additional constraints on the orthogonal mate,
they can be checked at this point.  Alternatively, it may be possible
to run the Euler--Parker algorithm in such a way that it will \emph{always}
construct orthogonal mates that satisfy the additional constraints.
For example, in Section~\ref{sec:hybrid_myrvold} we show how to do this for Myrvold's $\MOLS{2}{10}$ instances.

When the values in square $P$ have been completely assigned to form a square $L$ which
has been determined to have no orthogonal mates, the blocking clause passed to the solver is
\[ \bigvee_{L[i,j]=k} \lnot P_{i,j,k} \]
which naively contains $n^2$ literals, since $i$ and $j$ both range over $\{0,\dotsc,n-1\}$.
In fact, it is sufficient to take $i$ and $j$ only over $\{0,\dotsc,n-2\}$, since
the upper-left $(n-1)\times(n-1)$ subsquare of $L$ uniquely determines the final row and column.
This shrinks the length of the blocking clause to $(n-1)^2$ literals,
though this clause still only blocks the square~$L$.

\subsection{Specialization to Myrvold's $\MOLS{2}{10}$ instances} \label{sec:hybrid_myrvold}

The encoding described so far is for the general $\MOLS{2}{n}$ problem.
Our experimentation of this general encoding as $n$ grows
reveals the augmented SAT solver dramatically outperforms a pure SAT solver (see Section~\ref{sec:results}).
While this is a promising sign of the usefulness
of the hybrid approach, the general $\MOLS{2}{n}$ problem can be solved without
a SAT solver and so this is a somewhat artificial benchmark.

To highlight the utility of our hybrid approach, we use it to solve
$\MOLS{2}{10}$ instances with structure that seems intractable to handle purely
theoretically.  The eight $\MOLS{2}{10}$ cases left unsolved by Myrvold
(see Section~\ref{sec:background_myrv})
are suitable for this purpose, since she was unable to solve the eight cases
by theoretical arguments.  Currently, the cases have only 
been solved with the assistance of a SAT solver~\cite{BrightKeitaStevens2026Myrvold}.
In order to enforce the additional structure
considered by Myrvold, we
add more constraints on the pair $(P,Q)$ from Section~\ref{sec:encoding}.
For example, there are colour
constraints, subsquare $\Omega$ compatibility constraints, and symmetry breaking constraints.
There are also constraints stating that the pair $(P,Q)$ is one of Myrvold's eight unsolved
types.  For example, $(P,Q)$ being of type XX means that both $P$ and~$Q$ contain four
transversals of type $p_1$ and six transversals of type $p_2$.
A transversal is of type $p_i$ when it contains exactly $i$ white entries in the last
four columns, so the type constraints can be implemented by adding
appropriate cardinality constraints into the SAT instance.
Because these constraints are not an essential component of our
new hybrid approach, we do not go into them in more detail and instead refer the reader
to~\cite[Sec.~4]{BrightKeitaStevens2026Myrvold}.

The first stage of the Euler--Parker implementation does not need modification
to work on the instances encoding Myrvold's problem.
However, after the first stage, some transversals can be discarded because they
cannot be part of a valid solution.  For example, in the case of XX, we are looking
to find 10 disjoint transversals of the first square: 4 of type $p_1$ and 6 of type $p_2$.  So, any
transversals of type $p_3$ found by stage~1 will be discarded before starting stage~2.
Similarly, transversals that violate the $\Omega$ compatibility constraints
will also be discarded.
This filtering reduced the number of transversals in the average square
found by the SAT solver
from around 825 to around 140. Without loss of generality, there are two
possibilities for the $\Omega$ subsquare ($\Omega_1$ and $\Omega_2$),
and the SAT instance permits solutions
from either possibility---so
long as solutions only contain transversals all compatible with $\Omega_1$
or all compatible with $\Omega_2$. 
In order to enforce this in our hybrid approach, we run the second Euler--Parker stage twice,
once only with transversals compatible with $\Omega_1$, and once only with transversals
compatible with $\Omega_2$.

In the second stage of the Euler--Parker implementation we make use of
Myrvold's constraints explicitly.  For example, in case XX we specify that
a valid solution requires exactly four transversals of type $p_1$ and six transversals of type $p_2$.
Letting $T_i\subseteq T$ denote the set of transversals of type $p_i$, we form the equations
$\sum_{t\in T_1}x_t=4$ and $\sum_{t\in T_2}x_t=6$ and pass them to {\libexact}.
There are also colour constraints that can be directly incorporated:
a valid solution must possess exactly two
dark entries in each of the first six columns.  Let $D_i\subseteq T$ denote the transversals whose
entry in the $i$th column is coloured dark.  We form the six equations $\sum_{t\in D_i}x_t=2$
for $0\leq i<6$ and pass them all to the linear system provided to {\libexact}.

\section{Results}\label{sec:results}

In this section, we evaluate the performance of our hybrid SAT + Euler--Parker method.
As a first experiment, we test its performance on the unrestricted $\MOLS{2}{n}$
problem (see Section~\ref{sec:results_unrestricted}).
Our primary experiment is solving each of the eight $\MOLS{2}{10}$
cases that Myrvold left unsolved (see Section~\ref{sec:results_myrvold}).
All instances were solved using the SAT solver {\CaDiCaL} version 3.0, and the Euler--Parker
implementation used {\libexact} version 1.0.
Experiments were run on the Digital Research Alliance of Canada's Fir cluster,
a high-performance computing cluster consisting of AMD EPYC processors,
most of which run at 2.7~GHz. Our source code is available at~\href{https://github.com/aaronbarnoff/MOLS\_EP}{https://github.com/aaronbarnoff/MOLS\_EP}.

\subsection{Results for $\MOLS{2}{n}$ instances} \label{sec:results_unrestricted}

As pointed out in Section~\ref{sec:intro}, the unrestricted $\MOLS{2}{n}$ problem
has been solved by purely theoretical constructions, and therefore one does not
need to use a SAT solver for this problem.  Even still, it forms a useful
experiment to demonstrate that providing SAT solvers
with static Euler--Parker constraints (i.e., constraints asserting
the symbols of one square decompose the other square into $n$ disjoint transversals)
may not be as effective as running an external Euler--Parker implementation.

Table~\ref{tbl:2molsn} provides a detailed timing breakdown of our
unrestricted $\MOLS{2}{n}$ results for $8\leq n\leq 12$.
The table shows that as $n$ grows the hybrid SAT + Euler--Parker approach
is orders of magnitude faster than a standard SAT solver provided with static Euler--Parker
constraints.  For example, for $n=12$, the median time across 15 tests for
the hybrid approach is at least 150,000 times faster than the median time for the pure SAT
approach (for which the solver did not finish within four days).

\begin{table}[ht]
\caption{Runtime breakdown for the $\MOLS{2}{n}$ experiments, showing median times in seconds
for the pure SAT and the hybrid SAT + Euler--Parker approach across 15 random instances.
The median number of Euler--Parker calls before an orthogonal mate was found is also provided.
The timeout was four days.
}
\centering
\begin{tabular}{
c
S[table-format=1.2,group-digits=true]
S[table-format=3.2,group-digits=true]
S[table-format=4.2,group-digits=true]
S[table-format=5.2,group-digits=true]
S[table-format=1.2,group-digits=true]
}
{Order $n$} & {8} & {9} & {10} & {11} & {12} \\
\hline
Pure SAT     & 0.09 & 168.19  & 3486.03  & 39383.54   & \tout \\
Hybrid       & 0.26 & 0.42    & 0.78    & 1.20      & 2.23  \\
\# EP calls  & {160} & {31} & {2} & {1} & {1} \\
\end{tabular}
\label{tbl:2molsn}
\end{table}

These results conclusively demonstrate that off-the-shelf SAT solvers
do not exploit the Euler--Parker approach to its full potential.
As $n$ increased, our experimentation showed that the static Euler--Parker
constraints made the SAT solver find a square with an orthogonal mate quickly.  For
$n\geq11$, in the median case the \emph{very first square} the solver found
was one having an orthogonal mate.  Despite this, the pure SAT solver
was unable to find its mate, evidently backtracking
before finding $n$ disjoint transversals in the square.
Of course, the pure SAT solver is only provided static constraints
telling it to find $n$ disjoint transversals; it does
not attempt to find \emph{all} transversals
and then from within these try to find $n$ disjoint transversals.
This two-stage methodology is exploited by the hybrid approach
and explains why it can dramatically outperform the pure SAT approach.

These instances had {\CaDiCaL}'s \verb|--shufflerandom| option enabled, which randomizes the initial variable scores and causes the solver
to branch on different variables in each experiment.  We noticed that
otherwise the hybrid solver would always make the same decisions
and find the same solutions every time, even when run using different random seeds.
This seems to be an artifact of the fact that the hybrid solver is so fast
it solves the instances before the impact of the random seed is felt.
The hybrid solver also had {\CaDiCaL}'s \verb|--preprocesslight| disabled
because otherwise in some orders the solver would find a solution instantly
as a result of the preprocessing, leading to the same decisions being made
every time, even with variable shuffling enabled.

\subsection{Results for Myrvold's $\MOLS{2}{10}$ instances} \label{sec:results_myrvold}

\begin{figure}
    \centering
    \includegraphics[width=0.95\linewidth]{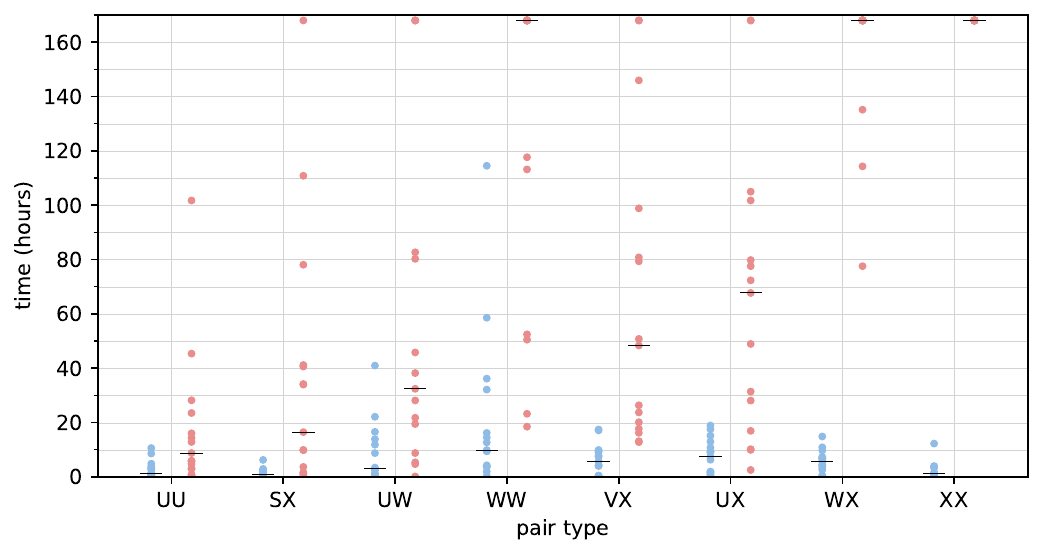}
    \caption{Scatterplot comparing the hybrid approach (left) to the pure SAT approach (right), for the eight Myrvold pair types. The median time is indicated with a black bar.
    }
    \label{fig:myrv_scatterplot}
\end{figure}

We now describe the results of our hybrid SAT + Euler--Parker approach on Myrvold's eight unresolved cases,
and compare them with the pure SAT approach.
For both approaches, we ran 15 instances of each of the eight cases for seven days,
using a different random seed for each instance so that {\CaDiCaL} would make different decisions each time.
These results are summarized in Table~\ref{tab:myrv_results_combined},
with a visual comparison shown in Figure~\ref{fig:myrv_scatterplot}.

\begin{table}[ht]
\caption{Comparison of the hybrid and pure-SAT approaches for each of the eight Myrvold pair types. Times are in seconds. Each pair type had 15 randomly seeded instances that ran for one week.}
\centering
\begin{tabular}{
c
c
c
S[table-format=6.2,group-digits=true]
S[table-format=6.2,group-digits=true]
S[table-format=6.2,group-digits=true]
}
\hline
pair type & method & {\# solved} & {median} & {min} & {max} \\
\hline
\multirow{2}{*}{UU} & Hybrid   & 15 & 5077.36   & 509.92    & 38323.47 \\
                    & Pure SAT & 15 & 31721.01  & 2706.00   & 366291.48 \\
\hline
\multirow{2}{*}{SX} & Hybrid   & 15 & 2882.27   & 499.79    & 22652.56 \\
                    & Pure SAT & 14 & 59447.47  & 255.08    & {\tout} \\
\hline
\multirow{2}{*}{UW} & Hybrid   & 15 & 11304.32  & 903.01    & 147568.43 \\
                    & Pure SAT & 12 & 116856.78 & 751.59    & {\tout} \\
\hline
\multirow{2}{*}{WW} & Hybrid   & 15 & 35638.79  & 1360.37   & 412264.03 \\
                    & Pure SAT & 6  & {\tout}   & 66699.78  & {\tout} \\
\hline
\multirow{2}{*}{VX} & Hybrid   & 15 & 20789.28  & 484.36    & 62917.17 \\
                    & Pure SAT & 13 & 174215.01 & 46195.26  & {\tout} \\
\hline
\multirow{2}{*}{UX} & Hybrid   & 15 & 27033.18  & 205.98    & 68228.84 \\
                    & Pure SAT & 13 & 243781.81 & 9314.35   & {\tout} \\
\hline
\multirow{2}{*}{WX} & Hybrid   & 15 & 20612.80  & 491.44    & 53664.08 \\
                    & Pure SAT & 3  & {\tout}   & 279253.68 & {\tout} \\
\hline
\multirow{2}{*}{XX} & Hybrid   & 15 & 5105.64   & 491.25    & 44225.28 \\
                    & Pure SAT & 0  & {\tout}   & {\tout}   & {\tout} \\
\hline
\end{tabular}
\label{tab:myrv_results_combined}
\end{table}

We found that the best performance was obtained by limiting how often a call to Euler--Parker could be invoked.
This was due to how quickly {\CaDiCaL} produces completed squares, each of which makes a relatively expensive call to {\libexact} in the first exact cover stage.
When the call to Euler--Parker was unrestricted, it was called so often that the SAT solver 
had little opportunity to make progress of its own through conflict-driven search,
as reflected in a low number of internal (non-programmatic) conflicts and restarts. 
The blocking clauses generated by Euler--Parker were added to the solver as forgettable
clauses (meaning that the solver was allowed to discard them to keep memory usage low), but {\CaDiCaL}
only performs reductions after a certain number of non-programmatic conflicts.
Since programmatic blocking clauses were being created so rapidly,
memory usage increased steadily and typically exceeded 8~GiB after running for a few days. 

To address this, we configured the solver so that the solver generates at least one
non-programmatic conflict clause before calling Euler--Parker again.
This threshold gave the SAT solver more time to progress through the search space between Euler--Parker invocations.
This resulted in far fewer squares being passed to Euler--Parker between clause database reductions, substantially reducing memory usage.
The highest recorded memory usage by the hybrid solver was 0.99~GiB (after running for about 5 days). We also found that the threshold improved the performance of the solver in most cases, particularly UW (see Table~\ref{tab:myrv_no_threshold}).
A breakdown of the median time spent in the SAT solver and in each stage of the Euler--Parker algorithm is shown in Table~\ref{tab:myrv_breakdown}.

\begin{table}[ht]
\caption{Comparison of the hybrid method in solving the eight Myrvold cases with and without limiting the frequency of Euler--Parker (EP) calls. Time is median solve time in seconds across 15 randomly seeded instances.
}
\centering
\setlength{\tabcolsep}{4pt}
\begin{tabular}{
c
S[table-format=5.2,group-digits=true]
S[table-format=5.2,group-digits=true]
S[table-format=6.2,group-digits=true]
S[table-format=5.2,group-digits=true]
S[table-format=5.2,group-digits=true]
S[table-format=5.2,group-digits=true]
S[table-format=5.2,group-digits=true]
S[table-format=4.2,group-digits=true]
}
\hline
{pair type} & {UU} & {SX} & {UW} & {WW} & {VX} & {UX} & {WX} & {XX} \\
\hline
Limited EP & 5077.36 & 2882.27 & 11304.32 & 35638.79 & 20789.28 & 27033.18 & 20612.80 & 5105.64 \\
Unlimited EP & 20871.53 & 27065.39 & 230812.17 & 87203.51 & 38492.24 & 18606.29 & 14058.23 & 8672.98 \\
\hline
\end{tabular}
\label{tab:myrv_no_threshold}
\end{table}

\begin{table}[ht]
\caption{Runtime breakdown for the eight Myrvold cases, showing median times in seconds before a $\MOLS{2}{10}$ was found.
The total runtime is divided into the SAT solving time and the time for the two stages of the Euler--Parker (EP) method.
The median number of calls to Euler--Parker is also provided.
}
\centering
\setlength{\tabcolsep}{4pt}
\begin{tabular}{
c
S[table-format=4.2,group-digits=true]
S[table-format=5.2,group-digits=true]
S[table-format=5.2,group-digits=true]
S[table-format=5.2,group-digits=true]
S[table-format=5.2,group-digits=true]
S[table-format=5.2,group-digits=true]
S[table-format=5.2,group-digits=true]
S[table-format=4.2,group-digits=true]
}
\hline
{pair type} & {UU} & {SX} & {UW} & {WW} & {VX} & {UX} & {WX} & {XX} \\
\hline
SAT solver & 2036.67 & 960.89 & 6788.22 & 12107.96 & 6847.76 & 8957.87 & 5366.88 & 1095.49 \\
EP stage 1 & 2698.06 & 1753.94 & 3929.43 & 20634.07 & 12179.40 & 15456.45 & 12957.77 & 3274.35 \\
EP stage 2 & 372.91 & 173.16 & 586.67 & 3116.00 & 1762.12 & 2618.85 & 2288.15 & 625.46 \\
\# EP calls & {1\,253\,736} & {677\,695} & {1\,565\,150} & {8\,148\,047} & {4\,688\,434} & {5\,974\,377} & {4\,928\,770} & {1\,254\,684} \\

\hline
\end{tabular}
\label{tab:myrv_breakdown}
\end{table}

For these instances, we call Euler--Parker whenever the SAT solver finds
a complete assignment of either the variables defining the square~$P$
or the square~$Q$.  When~$P$ has been determined, we run Euler--Parker
to construct~$Q$, and when~$Q$ has been determined, we run Euler--Parker
to construct~$P$.
Without loss of generality, the first row of $P$ can be taken to be one of
three possibilities~\cite[Thm.~13]{BrightKeitaStevens2026Myrvold},
and the SAT encoding enforces the first row of $P$ to be one of these three possibilities.
In cases when $Q$ has been determined, it is also possible to add constraints
into the second {\libexact} instance enforcing that one of the three possibilities
appears in the transversal decomposition of $Q$.
However, it was more efficient to \emph{not} provide this
symmetry breaking constraint to {\libexact}, thereby making
the stage~2 instance less constrained.
The downside is that
it makes the {\libexact} call less efficient.  However,
in general it results in fewer calls to Euler--Parker, as
it allows the solver to find solutions that it otherwise would discard.

\section{Conclusion} \label{sec:conclusion}

In this paper, we proposed and implemented a novel strategy for constructing orthogonal Latin squares
by incorporating both a SAT solver and the Euler--Parker strategy for orthogonal mate
construction.  Our approach exploits both the powerful search-with-learning capabilities
of a modern SAT solver, as well as the rich mathematical structure present in orthogonal
Latin squares.  We use our new strategy
to solve a number of challenging orthogonal Latin square problems arising
in the work of Myrvold~\cite{Myrvold1999NegativeResults}.  We stress that our method crucially exploits
\emph{both} a SAT solver and
the mathematical structure used by Euler~\cite{euler1782recherches} and Parker~\cite{parker1959computer}.
The SAT solver searches for an initial
starting square with an orthogonal mate having structure concisely expressed
in conjunctive normal form.  Conversely, the
Euler--Parker algorithm is used to efficiently construct an orthogonal mate from a
starting square.  Note that the starting square needs to be found before Euler--Parker can be used, and
in many cases, including in the work of Myrvold, it is not clear how to find a
square having an orthogonal mate with the proper structure.

As a case study, we build on the work of
Bright, Keita, and Stevens~\cite{BrightKeitaStevens2026Myrvold}, who recently
reduced to SAT eight $\MOLS{2}{10}$ cases left open by Myrvold.
The reduction to SAT was key to resolving
these cases, as they had been left unsolved for over 25 years.
Despite the success in using an off-the-shelf SAT solver to resolve these cases,
we show our hybrid SAT + Euler--Parker approach
dramatically outperforms the pure SAT approach.  On the hardest of the eight
cases, the hybrid solver typically takes 1.4 hours to solve what the pure SAT approach
does not solve in a week.
We believe our approach will
also be useful in searching for other kinds of orthogonal Latin squares, and this
will be the subject of future work.

\section*{Acknowledgements}

We thank the referees for several useful comments which improved the paper.

\section*{Declaration on Generative AI}

During the preparation of this work, the authors used ChatGPT and Gemini for proofreading. Additionally, A. Barnoff used OpenAI Codex to review C++ code for possible correctness issues and potential implementation optimizations.
The authors reviewed and edited the content as needed and take full responsibility for the publication's content.

\bibliography{paper}

\end{document}